\documentclass[12pt]{article}
\usepackage[dvips]{graphicx}

\usepackage{epsfig,amsmath,amssymb,verbatim,mathrsfs}

\def\beq{\begin{eqnarray}}
\def\eeq{\end{eqnarray}}
\def\bea{\begin{eqnarray}}
\def\eea{\end{eqnarray}}

\def\tev{\, {\rm TeV}}
\def\gev{\, {\rm GeV}}
\def\mev{\, {\rm MeV}}

\newcommand{\gsim}{\lower.7ex\hbox{$\;\stackrel{\textstyle>}{\sim}\;$}}
\newcommand{\lsim}{\lower.7ex\hbox{$\;\stackrel{\textstyle<}{\sim}\;$}}

\def\stilde{\widetilde}

\newcommand{\newc}{\newcommand}
\newc{\Nc}{N_{c}}
\newc{\CG}{C_G}
\newc{\gp}{g'}
\newc{\stopi}{\stilde t_i}
\newc{\sboti}{\stilde b_i}
\newc{\staui}{\stilde \tau_i}
\newc{\stopj}{\stilde t_j}
\newc{\sbotj}{\stilde b_j}
\newc{\stauj}{\stilde \tau_j}
\newc{\stopI}{\stilde t_1}
\newc{\stopII}{\stilde t_2}
\newc{\sbotI}{\stilde b_1}
\newc{\sbotII}{\stilde b_2}
\newc{\stauI}{\stilde \tau_1}
\newc{\stauII}{\stilde \tau_2}
\newc{\sstop}{s_{t}}
\newc{\cstop}{c_{t}}
\newc{\ssbot}{s_{b}}
\newc{\csbot}{c_{b}}
\newc{\sstau}{s_{\tau}}
\newc{\cstau}{c_{\tau}}
\newc{\Sstop}{s_{2t}}
\newc{\Cstop}{c_{2t}}
\newc{\Ssbot}{s_{2b}}
\newc{\Csbot}{c_{2b}}
\newc{\Sstau}{s_{2\tau}}
\newc{\Cstau}{c_{2\tau}}
\newc{\salpha}{s_\alpha}
\newc{\calpha}{c_\alpha}
\newc{\Calpha}{c_{2\alpha}}
\newc{\Salpha}{s_{2\alpha}}
\newc{\sbetapm}{s_{\beta_\pm}}
\newc{\cbetapm}{c_{\beta_\pm}}
\newc{\Sbetapm}{s_{2 \beta_\pm}}
\newc{\Cbetapm}{c_{2 \beta_\pm}}
\newc{\sbetaO}{s_{\beta_0}}
\newc{\cbetaO}{c_{\beta_0}}
\newc{\SbetaO}{s_{2 \beta_0}}
\newc{\CbetaO}{c_{2 \beta_0}}
\newc{\vu}{v_u}
\newc{\vd}{v_d}
\newc{\seL}{\stilde e_L}
\newc{\smuL}{\stilde \mu_L}
\newc{\seR}{\stilde e_R}
\newc{\smuR}{\stilde \mu_R}
\newc{\suL}{\stilde u_L}
\newc{\sdL}{\stilde d_L}
\newc{\suR}{\stilde u_R}
\newc{\sdR}{\stilde d_R}
\newc{\scL}{\stilde c_L}
\newc{\ssL}{\stilde s_L}
\newc{\scR}{\stilde c_R}
\newc{\ssR}{\stilde s_R}
\newc{\snue}{\stilde \nu_e}
\newc{\snumu}{\stilde \nu_\mu}
\newc{\snutau}{\stilde \nu_\tau}
\newc{\Gpm}{G^\pm}
\newc{\Hpm}{H^\pm}
\newc{\FFbS}{\overline{FF}S}
\newc{\FFbV}{\overline{FF}V}
\newc{\FSS}{F_{SS}}
\newc{\FSSS}{F_{SSS}}
\newc{\FFFS}{F_{FFS}}
\newc{\FFFbS}{F_{\overline{FF}S}}
\newc{\FSSV}{F_{SSV}}
\newc{\FVS}{F_{VS}}
\newc{\FVVS}{F_{VVS}}
\newc{\FFFV}{F_{FFV}}
\newc{\FFFbV}{F_{\overline{FF}V}}
\newc{\Fgauge}{F_{\rm gauge}}
\newc{\DRbarprime}{$\overline{\rm DR}'$ }
\newc{\DRbar}{$\overline{\rm DR}$ }
\newc{\MSbar}{$\overline{\rm MS}$ }
\newc{\Yu}{{\bf Y}_u}
\newc{\Yd}{{\bf Y}_d}
\newc{\Ye}{{\bf Y}_e}
\newc{\Au}{{\bf a}_u}
\newc{\Ad}{{\bf a}_d}
\newc{\Ae}{{\bf a}_e}
\newc{\bm}{{\bf m}}
\newc{\zhol}{Z^{\rm hol}}

\newc{\rwino}{r_{\tilde W}}
\newc{\rmu}{r_{\tilde H}}
\newc{\ra}{r_A}
\newc{\ccdot}{\!\cdot\!}

\newcommand{\nnmb}{\nonumber}
\newcommand{\del}{\partial}

\newcommand{\lrf}[2]{\left(\frac{#1}{#2}\right)}

\newcommand{\kev}{\mbox{keV}}

\begin{document}

\setlength{\baselineskip}{0.2in}


\begin{titlepage}
\noindent
\vspace{1cm}

\begin{center}
  \begin{Large}
    \begin{bf}
Low-Energy Probes of a Warped Extra Dimension
     \end{bf}
  \end{Large}
\end{center}
\vspace{0.2cm}

\begin{center}

\begin{large}
Kristian L. McDonald and David E. Morrissey\\
\end{large}
\vspace{0.3cm}
  \begin{it}
TRIUMF\\
4004 Wesbrook Mall, Vancouver, BC V6T 2A3, Canada\vspace{0.5cm}\\
\end{it}

\end{center}

\center{September 21, 2010}

\begin{abstract}

We investigate a natural realization of a light Abelian hidden sector 
in an extended Randall-Sundrum (RS) model.  
In addition to the usual RS bulk we consider a second warped space 
containing a bulk $U(1)_x$ gauge theory with a 
characteristic IR scale of order a GeV. 
 This Abelian hidden 
sector can couple to the standard model via gauge kinetic mixing 
on a common UV brane.  We show that if such a coupling induces
significant mixing between the lightest $U(1)_x$ gauge mode and the
standard model photon and Z, it can also induce significant mixing 
with the heavier $U(1)_x$ Kaluza-Klein (KK) modes.  As a result 
it might be possible to probe several KK modes in upcoming fixed-target
experiments and meson factories, thereby offering a new way 
to investigate the structure of an extra spacetime dimension.  

\end{abstract}

\vspace{1cm}

\end{titlepage}

\setcounter{page}{2}


\vfill\eject



\section{Introduction}

New physics beyond the standard model (SM) should either be 
kinematically inaccessible or couple very weakly to the SM in order 
to have evaded experimental efforts to date.  The first possibility, 
of heavy new physics, has received 
the most attention and has pushed us to build ever 
larger colliders in an attempt
to directly probe new energy 
frontiers~\cite{Ball:2007zza,Aad:2009wy,Morrissey:2009tf}. 
It has also motivated the construction of precise lower-energy 
probes that search for the indirect effects 
of heavy new physics~\cite{Gabbiani:1996hi,Harrison:1998yr,:2005ema}.
The second possibility, of relatively light new 
physics which couples very weakly to the SM, has not been 
explored to the same extent.
Such new physics can also be sought in high-energy particle
colliders, but in many cases a more promising route is to use
lower-energy colliders with an enormous luminosity. 

 Hidden sectors containing new particles coupled
very weakly to the SM
comprise an interesting class of new physics 
scenarios that emerge naturally in
a number of
extensions of the SM. 
For example, theories of supersymmetry breaking typically contain a 
hidden sector where the breaking actually occurs. This hidden sector 
only couples to the SM via a set of heavy mediator 
particles~\cite{Martin:1997ns} and  
in some cases can give rise to very light 
states~\cite{Nelson:1993nf}.  
Several recently proposed theories of 
dark matter, motivated by astrophysical measurements such as  
PAMELA~\cite{Adriani:2008zr} and Fermi~\cite{Abdo:2009zk}, 
also contain light hidden sectors~\cite{ArkaniHamed:2008qn}.
New gauge groups under which the SM fields are singlets 
also arise frequently in string-theoretic 
constructions~\cite{Giddings:2001yu,Grana:2005jc}. 

  If the characteristic mass scale of a hidden sector is at 
or below a few GeV, it can potentially be discovered in current and upcoming 
fixed-target~\cite{Borodatchenkova:2005ct,Bjorken:2009mm,Batell:2009di,Freytsis:2009bh,Schuster:2009au} 
and $e^+e^-$ colliders~\cite{Pospelov:2008zw,Batell:2009yf,Essig:2009nc,Reece:2009un,Bossi:2009uw,Yin:2009mc}.
It is therefore timely and interesting to study hidden sector models 
where the GeV scale emerges in a (technically) natural way.  
Previous works have shown that the GeV scale can arise naturally in
supersymmetric models provided  supersymmetry breaking 
in the hidden sector is suppressed relative to the SM~\cite{
ArkaniHamed:2008qp}.  
Similarly, models with composite hidden states can naturally realize 
a light hidden sector via dimensional transmutation~\cite{Alves:2009nf},
and have been studied in the context of hidden valleys~\cite{Strassler:2006im,
Strassler:2008bv} and unparticles~\cite{Georgi:2007ek,Stephanov:2007ry}.

\begin{figure}[ttt]
\begin{center}
        \includegraphics[width = 0.5\textwidth]{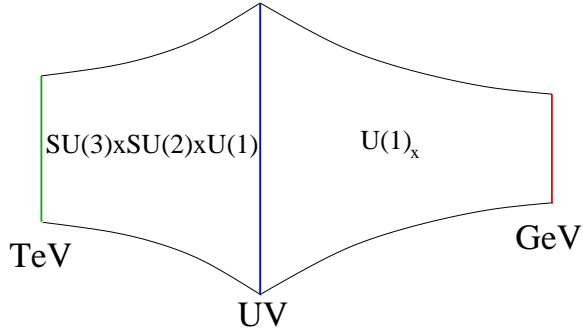}
\end{center}
\caption{Schematic diagram of the scenario under consideration.}
\label{rshidpic}
\end{figure}

 In this work we realize a light Abelian hidden 
sector in an extended Randall-Sundrum~(RS) model~\cite{Randall:1999ee}. 
RS models provide a geometric means by which to naturally generate mass
hierarchies and can readily realize sub-Planckian scales
as simple redshifted incarnates of Planck scale input parameters. 
The specific model we consider consists 
of the standard bulk RS scenario together with a second hidden bulk space. 
The SM propagates exclusively in one of the bulks with the Higgs localized 
on the IR brane to realize the weak/Planck hierarchy. The second hidden bulk 
shares the same UV brane but has an independent IR scale. We take the hidden 
IR scale to be near a GeV and, as a minimal scenario, 
consider a $U(1)_x$ gauge theory propagating in the hidden bulk.  
An illustration of the setup appears in Figure~\ref{rshidpic}.

At energies near a GeV the minimal spectrum consists of the SM along with 
towers of hidden GeV-spaced gauge and gravity KK modes. 
The SM couples to the hidden KK vectors primarily via a localized kinetic 
mixing operator connecting $U(1)_x$ and $U(1)_Y$ on the shared UV brane.  
This kinetic mixing induces a suppressed coupling between the hidden 
vectors and SM matter which, for the lightest hidden KK modes, 
is proportional to SM electric charge~\cite{Holdom:1985ag}. 
Relative to the lightest $U(1)_x$ mode,
the higher modes have a coupling that is suppressed, but only by
a moderate logarithmic factor.  This can potentially allow the 
creation of several hidden gauge KK modes at high luminosity
low-energy particle colliders.  Kinetic mixing also allows these modes to decay
back to the SM and we show that such channels can dominate, at
least for the first few KK states.  Fixed-target experiments and meson
factories therefore have the potential to explore the geometry of a warped 
extra dimension beyond what might
be possible at the LHC.

  Although there is nothing fundamental about our choice of a GeV for 
the hidden IR scale, we feel there is a strong phenomenological motivation 
to study models of this type. 
RS models can be thought of as effective theory frameworks that 
capture some of the gross features of the warped throats found
in string theory. The simplifications of the RS model include neglecting 
transverse throat structure\footnote{Phenomenological extensions
 that model  transverse structure are possible; 
see e.g.~\cite{Davoudiasl:2002wz}.} and shrinking the rest of the 
compactification manifold
to a single point (represented by the UV brane).
String compactifications can also contain
multiple warped
throats with 
distinct characteristic IR scales~\cite{Giddings:2001yu,Grana:2005jc,Hebecker:2006bn}.
This situation
can be modeled with 
multiple RS bulks sharing a common
UV brane~\cite{Cacciapaglia:2006tg}.
The essential point is that if the SM is localized in one throat
and has similar couplings
to  other throats, the throats
with the lowest IR scales are
typically the easiest to detect.

RS models are also thought to be dual to certain classes of 
4D conformal field theories~(CFTs) with conformal symmetry breaking
at energies corresponding to the UV and IR scales~\cite{ArkaniHamed:2000ds}.
In the multi-throat picture the hidden throats are dual to additional 
hidden 4D CFTs that share a common UV breaking of conformal invariance.
These conformal breaking interactions 
can induce couplings between the different CFTs which are modeled by UV
localized operators in the 5D picture. Each of the hidden CFTs has
an, in principle, independent IR scale at which the conformality of
the given CFT is broken. These IR scales set the characteristic
scale for low-energy composites of the CFTs. We shall use the language of
the 5D multi-throat picture in the present work but one should keep in
mind that our investigations also encompass potentially observable
effects from a broad class of hidden CFTs that couple to the SM.
As with multiple warped throats, in a theory with multiple hidden CFTs
we typically expect those with light 
characteristic mass scales to be the most amenable to discovery.

  Multiple warped bulk spaces have been considered previously in 
a number of works.   Our effective theory description follows that 
of Refs.~\cite{Cacciapaglia:2006tg,Agashe:2007jb}.  
Various aspects of multi-throat physics from a 
string-motivated perspective appear 
in Refs.~\cite{Hebecker:2006bn}, and phenomenological
applications can be found in Refs.~\cite{Dimopoulos:2001ui,
Cacciapaglia:2005pa,Flacke:2006ad}.
Kinetic mixing in warped spaces was investigated in Ref.~\cite{Batell:2005wa}
as well as in certain string compactifications in 
Refs.~\cite{Dienes:1996zr,Abel:2003ue}.
Our work also has overlap with several studies of 
hidden valleys~\cite{Strassler:2006im,Strassler:2008bv},
composite dark sectors~\cite{Alves:2009nf}, and 
unparticles~\cite{Strassler:2008bv,Georgi:2007ek,Stephanov:2007ry}.
However, relative to these earlier studies, our specific construction
and the application to low-energy probes is new to the best of our knowledge.

  The outline of this paper is as follows.  
In Section~\ref{model} we describe the gravitational background 
of our model.  Next, we derive the spectrum of physical vectors and their 
couplings to SM matter in Section~\ref{sec:gauge}. 
In Section~\ref{signals} we consider the constraints on
this scenario and discuss the prospects for new signals in upcoming 
fixed-target and meson factory
experiments.  Finally, Section~\ref{conc} 
is reserved for our conclusions.  Some technical details including 
the KK graviton spectrum and details of the gauge boson mass 
diagonalization and couplings appear 
in the Appendix.

\section{Gravitational Background and Modes \label{model}}

  The extended RS model that we consider consists of two independent 
warped bulk spaces, or \emph{throats}, sharing a common UV brane. 
We use the coordinates $z_i \in [k^{-1},\,R_i]$, $i=1,2$,  
for the extra dimension of the $i$-th throat and take the metric in 
the $i$-th throat to be
\beq
ds^2_i = \frac{1}{(kz_i)^2}(\eta_{\mu\nu}dx^{\mu}dx^{\nu} -
dz_i^2)= G_{MN}^i dx^{M}dx^{N},
\label{bulkmetric}
\eeq
where $k$ is a throat-independent curvature\footnote{The
  curvature may differ in the two throats but will be approximately
  equal provided the throat localized energy densities are dominated by a
  common bulk cosmological constant. Allowing non-hierarchical
  differences will not significantly modify our results.}.  The
common UV brane, with characteristic mass
scale $\sim k$, is located at $z_1=z_2=k^{-1}$ and IR branes with 
characteristic scales $R_i^{-1}$ reside at
$z_i=R_i$ . The IR scales are suppressed relative to the curvature, 
as is readily seen in the non-conformal 
coordinates,\footnote{These are defined by 
$y_i= k^{-1}\log(kz_i)$, $y_i\in[0,\pi r_i]$ with 
$r_i= (\pi k)^{-1}\log(kR_i)$.} 
in terms of which one has $R_i^{-1}=e^{-k\pi r_i}k \ll k$. 
When sourced by a bulk cosmological constant and appropriate brane 
tensions the metric of Eq.~\eqref{bulkmetric}
is a solution to the 5D Einstein equations~\cite{Cacciapaglia:2006tg}.

  The SM (hidden sector) resides in the throat $i=1$
($i=2$) which we refer to as the visible (hidden) throat. 
The standard bulk RS picture is employed in the visible throat with SM gauge
and matter fields propagating in the bulk and the SM Higgs localized at 
$z_1=R_1$. Consistency with precision electroweak bounds requires 
$R_1^{-1} \gtrsim \tev$ as per usual in RS 
models~\cite{Agashe:2003zs,Carena:2006bn}.
In principle the hidden throat could  
have a complicated gauge structure 
and could contain
matter charged under the hidden gauge symmetries. However, in order 
to determine the likely low-energy probes of a light hidden sector 
it suffices to consider a minimal gauge sector. We therefore assume that 
the hidden throat contains an Abelian $U(1)_x$ gauge group that is 
broken, either spontaneously by an explicit Higgs at $z_2=R_2$ or 
through Higgsless IR boundary 
conditions~\cite{Csaki:2003dt}.   
The mass scale on the hidden IR brane is taken to be  
$R_2^{-1} \lesssim \gev$
and the setup is illustrated in 
Figure~\ref{rshidpic}.

Interactions between the SM and the hidden sector
will be mediated by gravity and by local operators on the UV brane.
Among the set of local operators, we focus primarily on 
a UV-localized gauge kinetic mixing interaction
\beq
S_{UV} \supset -\frac{\epsilon_*}{2M_*}
\int_{UV} d^4x\sqrt{-g}g^{\mu\alpha}g^{\nu\beta}B_{\mu\nu}X_{\alpha\beta},
\eeq
where $B_{\mu\nu}$ is the 5D hypercharge gauge field strength,
$X_{\mu\nu}$ is the 5D $U(1)_x$ field strength and $M_*$
is the the bulk gravity scale.  
This interaction produces the only {significant}
renormalizable coupling between
the SM and hidden sector fields in the effective 4D theory and is 
responsible for the low-energy effects of interest in this work.
 
The graviton KK spectrum contains a zero mode that extends into both 
throats and reproduces Newtonian gravity in the effective 4D theory. 
Its coupling is universal and dictated by the effective 4D Planck mass,
\begin{eqnarray}
M_{Pl}^2&=&\sum_i \frac{M_*^3}{2k}
\left[1-\frac{1}{(kR_i)^2}\right]
\simeq \frac{M_*^3}{k}.
\label{4d_planck}
\end{eqnarray}
For $k\sim M_*$ one has $M_{Pl}\sim M_*$ and we also take $k/M_*\lesssim 0.1$
to ensure the gravitational description can be trusted. 
In addition to the zero mode there are massive KK gravitons which
split up naturally into a pair of KK towers; 
see Appendix~\ref{app:KK_gravity} for details.
The first (second) tower has spacings on the order of 
$R_2^{-1}\sim \gev$ ($R_1^{-1}\sim \tev$)
and is strongly localized towards the IR of the hidden (visible) throat.  
One can show that the GeV-spaced KK modes
couple extremely weakly to matter on the UV brane
and to the SM in the visible throat. These light KK modes are therefore 
experimentally viable, which should come as no surprise to 
those familiar with the viability of gravity in RS2~\cite{Randall:1999vf}; 
the KK graviton spectrum here is essentially the RS2 spectrum with the 
sub-TeV modes removed on one side and the sub-GeV modes removed on the other.
The GeV-scale gravitons do however couple significantly to fields 
in the hidden bulk with important effects on the phenomenology 
of the hidden vectors.

  Metric fluctuations in the throats also give rise to a pair of physical 
scalar modes (radions),
that only acquire mass once a stabilization mechanism is specified. 
We will not discuss radius stabilization in this work, but
expect
that it is straightforward to extend the Goldberger-Wise 
mechanism~\cite{Goldberger:1999uk} to stabilize both throats. 
The ratio $R_2/R_1\sim10^3$ translates
into the rather mild hierarchy of $r_1/r_2\simeq 0.8$ in the non-conformal 
coordinates, which is easily achieved with small differences in the input 
parameters.  Beyond the tuning required to fix the 4D cosmological constant, 
the setup is therefore devoid of fine tunings.

\section{Gauge Bosons and Kinetic Mixing\label{sec:gauge}}

  In the setup described above the SM is sequestered from the hidden sector, 
up to gravity and local operators on the common UV brane. For the most part, 
the detailed bulk realization of the SM is not important for the low-energy 
effects we are interested in.  It will likely require a custodially-extended 
electroweak gauge sector~\cite{Agashe:2003zs} and may include additional 
flavor symmetries~\cite{Csaki:2008zd}.  The one assumption we 
make is that the hypercharge gauge factor extends to the UV brane,
and that it is the only Abelian factor to do so. Custodially-extended RS models 
typically invoke Dirichlet boundary conditions on the UV brane, 
leaving $U(1)_Y$ as 
the only Abelian factor to extend to the UV~\cite{Agashe:2003zs}. 
Our assumption is therefore consistent with standard bulk RS constructs.  

In what follows we 
derive the spectrum of hidden KK vectors and determine their couplings 
to the SM.  As we consider the low-energy physics our results would be 
qualitatively the same if the SM was simply placed on the UV brane with 
an alternative mechanism like supersymmetry stabilizing 
the weak/Planck hierarchy~\cite{Pomarol:2000hp}.\footnote{
In the CFT picture this corresponds to the SM being fundamental states, 
external to a hidden CFT.}

\subsection{Gauge Boson KK Modes}

  On the hidden side we consider two cases, differentiated by the breaking
of the $U(1)_x$ gauge symmetry.  In the first case $U(1)_x$ is broken
by the vacuum expectation value~(VEV) of a hidden Higgs $H_x$ confined
to the hidden IR brane with 
$\langle H_x \rangle \ll R_2^{-1}$.  
The second case employs Higgsless breaking of $U(1)_x$ via a Dirichlet
boundary condition on the hidden IR brane~\cite{Csaki:2003dt}.  
Increasing $\langle H_x \rangle$ towards and above $R_2^{-1}$ smoothly 
interpolates between these two cases~\cite{Csaki:2003dt}.  
Neumann boundary conditions are employed for both the SM and the hidden
gauge fields on the UV brane, {and we treat the kinetic mixing
operator as a small perturbation}.  

  The full bulk action for the Abelian gauge factors in unitary gauge
with $B_5 = X_5 = 0$ is
\bea
S &\supset& 
-\frac{1}{4}\int d^4x\,dz_1\;\sqrt{G_1}\,G_1^{MA}G_1^{NB}B_{MN}B_{AB}\\
&&- \frac{1}{4}\int d^4x\,dz_2\;\sqrt{G_2}\,G_2^{MA}G_2^{NB}X_{MN}X_{AB}\\
&&- \frac{\epsilon_*}{2M_*}
\int_{UV}\!\!d^4x\;\sqrt{-g}\,g^{\mu\nu}g^{\alpha\beta}
B_{\mu\nu}X_{\alpha\beta}.
\eea
We assume perturbative values of the mixing parameter $\epsilon_*\lesssim 1$, 
which could be generated by integrating out a set of UV localized Dirac
fermions charged under both hypercharge and $U(1)_x$. Regardless of its origin 
this term is consistent with the gauge symmetries of the 
theory and we therefore include it in our effective theory description. 

  Decomposing the five-dimensional $U(1)_x$ gauge field 
into KK modes according to\footnote{We drop the subscript in this section so 
$z=z_2$.}
\beq
X_{\mu}(x,z) = \sum_nf^{(n)}(z)\tilde{X}_{\mu}(x),
\eeq
the bulk wave functions satisfy the equation of motion
\beq
\left[z^2\partial_z^2 -z\partial_z + z^2m^2_n\right]f^{(n)}(z) = 0,
\eeq
and the following orthogonality conditions
\begin{eqnarray}
\int  \frac{dz}{(kz)}f^{(n)}(z)f^{(m)}(z)&=&\delta^{nm}.
\end{eqnarray}
In the weakly Higgsed case there is a zero mode with a
constant profile $f^{(0)} = \sqrt{k/\log(kR_2)}$.  Above this mode,
for both the Higgses and Higgsless cases, we have the wavefunctions
\beq
f^{(n)}(z) = \frac{(kz)}{N_n}\left[J_1(m_nz) +\beta_nY_1(m_nz)\right], 
\label{wavefunc-gauge}
\eeq
where $J_1$ and $Y_1$ are Bessel functions, 
and the Neumann boundary condition at $z=k^{-1}$ gives
\beq
\beta_n = -\frac{J_0(m_n/k)}{Y_0(m_n/k)} \simeq 
\frac{\pi}{2}\;\frac{1}{\log(2k/m_n)-\gamma},
\eeq
where $\gamma \simeq 0.5778$ 
is the Euler-Mascheroni constant.
The eigenvalues $m_n$ are fixed by applying the IR brane
boundary conditions, and give
\beq
\beta_n=-\frac{J_0(m_n R_2)}{Y_0(m_n R_2)} ~~~~~~~\mbox{
(Neumann at $z=R_2$)},
\label{neumannfreq}
\eeq
or
\beq
\beta_n=-\frac{J_1(m_n R_2)}{Y_1(m_n R_2)} ~~~~~~~\mbox{
(Dirichlet at $z=R_2$)}.
\label{dirichletfreq}
\eeq

 For $n$ greater than a few the KK masses $m_n$ and normalization factors 
$N_n$ are well approximated by 
\beq
m_n \simeq \frac{\pi}{R_2}(n\mp1/4),
\eeq
and
\beq
N_n^{-1} \simeq \frac{1}{(n\mp 1/4)^{1/2}}\frac{m_n}{\sqrt{k}},
\eeq
where the minus (plus) sign corresponds to the Higgsed (Higgsless) case.
The mass of the lowest mode in the Higgsless case,
which we label as ``0'', is suppressed relative to the hidden IR scale,
\beq
m_0 \simeq \frac{1}{R_2}\sqrt{\frac{2}{\log(2kR_2)-\gamma}},
\eeq
while its wavefunction is given by Eq.~\eqref{wavefunc-gauge} with
$N_0^{-1} \simeq \sqrt{2/kR_2^2}$.

Putting these results together, and treating the $\gamma$ and $Z$
components of the SM hypercharge gauge boson as simple zero modes with 
profiles $\sqrt{k/\log(kR_1)}$, the effective kinetic mixing operator is
\beq
\mathscr{L}_{eff} \supset -\frac{1}{2}\sum_{n}\epsilon_nX^{\mu\nu}_n
(c_W\tilde{F}_{\mu\nu}-s_W\tilde{Z}_{\mu\nu}),
\eeq
where $s_W$ and $c_W$ refer to the weak mixing angle, 
and the kinetic mixing parameters $\epsilon_n$ are given by
\beq
\epsilon_n = \epsilon_*\frac{k}{M_*}\frac{1}{\sqrt{\log(kR_1)}}
\frac{f^{(n)}(z=k^{-1})}{\sqrt{k}}.
\eeq

  For the zero modes, the expressions derived above 
in the weakly Higgsed scenario give 
\beq
\epsilon_0 \simeq 
{\epsilon_*}\frac{k}{M_*}\frac{1}{\sqrt{\log(kR_1)\log(kR_2)}},
~~~~~~~~~~~\text{(Higgsed)}
\eeq
while for the Higgsless case one has
\beq
\epsilon_0 \simeq 
{\epsilon_*}\frac{k}{M_*}\frac{1}{\sqrt{\log(kR_1)\,[\log(m_0/2k)+\gamma]}},
~~~~\text{(Higgsless)}.
\eeq
For the higher modes we find
\beq
\epsilon_n \simeq 
{\epsilon_*}\frac{k}{M_*}\frac{1}{\sqrt{\log(kR_1)}}\,
\frac{1}{[\log(m_n/2k)+\gamma]}(n\mp1/4)^{-1/2},
~~~{n \geq 1},
\label{epsilonn}
\eeq
where the minus (plus) sign again applies to the Higgsed (Higgsless) case.
These expressions show that the higher KK modes have a suppressed
kinetic mixing with the SM fields relative to the lowest mode, 
but that this suppression is only logarithmic.

\subsection{Neutral Gauge Boson Mass and Kinetic Mixing}

In the preceding we have motivated a tower of neutral vectors that 
kinetically mix with the SM in an extended RS framework. Having determined 
the mass spectrum and kinetic mixing parameters of the hidden vectors in 
the effective 4D theory we now present the transformations that diagonalize 
both the kinetic and mass mixing. These operations induce couplings between 
SM matter and the hidden vectors and also modify the couplings of the SM 
$Z$ boson, thereby providing experimental means by which to explore 
the present scenario. We note that despite having motivated the tower 
of spin-one modes via a warped extra dimension, the methodology we develop 
in this section is more general and can be employed in any scenario 
containing a tower of vectors that kinetically mix with SM hypercharge. 
We therefore keep the discussion somewhat general. For the time being we shall 
treat all KK modes as narrow resonances and will return to this point 
in Section~\ref{signals}.
    
We consider a tower of vectors $\tilde{X}_n$ with mass $m_n$ that kinetically 
mix with SM hypercharge. The tower is labeled by the integer 
$n\in [0,n_{\Lambda}]$ and the heaviest mode has 
mass $m_{n_{\Lambda}}\sim\Lambda$ where  $\Lambda\gg m_Z$ is a high-energy 
cutoff. In the present context we are primarily interested in the low-energy 
physics and do not concern ourselves with the KK modes of the SM photon 
and $Z$. We therefore truncate the hidden KK tower at the usual RS KK scale, 
$\Lambda\sim $~few TeV. For convenience we define the integer $n_z$ such that 
\begin{eqnarray}
m_n&<&m_Z \quad\mathrm{for }\quad n\le n_z,\\
m_n&>&m_Z \quad\mathrm{for }\quad n_z<n\le n_{\Lambda},
\end{eqnarray} 
where $m_Z$ is the SM value of the $Z$ mass. The mixed kinetic Lagrangian is
\begin{eqnarray}
-4\mathscr{L}_{Kin}&=&\tilde{F}_{\mu\nu}\tilde{F}^{\mu\nu}+\tilde{Z}_{\mu\nu}\tilde{Z}^{\mu\nu}+\sum_{n=0}^{n_{\Lambda}}\tilde{X}^n_{\mu\nu}\tilde{X}^{\mu\nu}_n+\sum_{n=0}^{n_{\Lambda}}2\epsilon_n(c_W \tilde{F}_{\mu\nu}-s_W \tilde{Z}_{\mu\nu})\tilde{X}^{\mu\nu}_n,\label{mixed_lagrangian_tower_e3}
\end{eqnarray}
where we label the mixed fields with a tilde and 
$(c_W,s_W)=(\cos\theta_W,\sin\theta_W)$ refer to the weak mixing angle.

In a general theory containing two vectors $V_{1,2}$ with masses 
$m_1< m_2$ and kinetic mixing parameter $\epsilon\ll 1$ the kinetic 
Lagrangian may be diagonalized to $\mathcal{O}(\epsilon^3)$ with the 
following field redefinitions:
\begin{eqnarray}
V_1&\rightarrow& V_1 -\epsilon V_2,\nonumber\\
V_2&\rightarrow& (1+\epsilon^2/2) V_2.
\end{eqnarray}  
This shift is asymmetric between $V_1$ and $V_2$
and the diagonalization can similarly be achieved by instead
shifting the heavier field $V_2$. However, by shifting the lighter
field (and simply rescaling the heavier one)
the mass mixing induced by the shift is proportional to the 
lighter mass scale $m_1$. Consequently the mass mixing angle is 
suppressed relative to the kinetic mixing parameter, $\sim
\epsilon m_1^2/m_2^2$, and in this basis mass mixing 
effects are subdominant. 

  With this in mind the strategy for decoupling 
the kinetic mixing in Eq.~\eqref{mixed_lagrangian_tower_e3}
is to always shift the lightest fields 
and thereby  minimize the effects of mass mixing. One first performs a 
shift of the photon field to decouple the kinetic mixing between 
$\tilde{A}$ and all the $\tilde{X}_n$. Then one decouples the mixing 
between $\tilde{Z}$ and $\tilde{X}_n$ by shifting $\tilde{X}_n$ for $n\le n_z$ 
and shifting $\tilde{Z}$ for $n>n_z$. The result of this multi-step shift 
can be combined into the following:
\begin{eqnarray}
\tilde{Z}^\mu&\rightarrow& \left(1+\frac{s_W^2}{2}\sum_{n=0}^{n_z}\epsilon_n^2\right)\tilde{Z}^\mu+s_W\sum_{n=n_z+1}^{n_\Lambda}\epsilon_n \tilde{X}_n^\mu,\nonumber\\
\tilde{X}^\mu_n&\rightarrow& \left\{\begin{array}{lcc}\left(1+c_W^2\epsilon_n^2/2\right)\tilde{X}^\mu_n+ \epsilon_n s_W\tilde{Z}^\mu&& n\le n_z\\
\left(1+\epsilon_n^2/2\right)\tilde{X}^\mu_n&& n> n_z
\end{array}\right.\label{e2_photon_shift_extended},\\
\tilde{A}^\mu&\rightarrow& A^\mu-\sum_{n=0}^{n_\Lambda} \epsilon_n c_W \tilde{X}^\mu_n - c_Ws_W\left(\sum_{n=0}^{n_z}\epsilon_n^2\right)\tilde{Z}^\mu.\nonumber
\end{eqnarray}
These shifts diagonalize the kinetic terms for $\tilde{Z}$ and $A$ up to 
$\mathcal{O}(\epsilon^3)$.  Subleading corrections of the form 
$\epsilon_n\epsilon_m\tilde{X}_n\tilde{X}_m$ remain but can be 
consistently neglected for $|\epsilon_n| \ll 1$.

 Performing the field redefinitions of Eq.~\eqref{e2_photon_shift_extended} 
in the vector-mass Lagrangian induces mass
mixing between $\tilde{Z}$ and $\tilde{X}_n$, while $A$ 
is the physical massless photon.  The full $\tilde{Z}$-$\tilde{X}_n$ 
mass matrix is given in Appendix~\ref{app:vec_mass_mixing}, 
and can be diagonalized by a series orthogonal transformations.  
To leading non-trivial order in $\epsilon$ 
(meaning $\mathcal{O}(\epsilon)$ in off-diagonal terms 
and $\mathcal{O}(\epsilon^2)$ on the diagonal),
a single independent orthogonal rotation is needed for each KK level.
The corresponding mixing angle at level $n$ is
\begin{eqnarray}
\eta_n&=&\epsilon_ns_W\times\frac{ m_<^2}{m_Z^2- m_n^2},
\end{eqnarray}
where
\begin{eqnarray}
m_<=\left\{\begin{array}{lcc}m_n&\quad\mathrm{for}&\quad n\le n_z\\
m_Z&\quad\mathrm{for}&\quad n> n_z
\end{array}\right..\label{m_<}
\end{eqnarray}
To $\mathcal{O}(\epsilon^2)$ the mixed fields $\tilde{Z}$, $\tilde{X}_n$ 
are related to the mass eigenstates $Z$, $X_n$ as
\begin{eqnarray}
\tilde{Z}^\mu&\simeq&
\left(1-\sum_{n=0}^{n_{\Lambda}}\frac{\eta_n^2}{2}\right)Z^\mu
-\sum_{n=0}^{n_{\Lambda}}\eta_nX^\mu_n,\\
\tilde{X}^\mu_n&\simeq&\left(1-\frac{\eta_n^2}{2}\right)X^\mu_n +\eta_nZ^\mu.
\end{eqnarray}
We give the corresponding mass eigenvalues in 
Appendix~\ref{app:vec_mass_mixing}.
With the above one readily obtains the coupling of the physical
vectors $X_n$ to SM matter and the induced modification of the SM
$Z$ coupling. 
We present these couplings in Appendix~\ref{cov_der_app}.

We note the following features of the above. For $n\ll n_z$ one has $m_n^2\ll m_Z^2$ and the 
mixing angle $\eta_n$ is mass-suppressed relative 
to the kinetic mixing parameter, $|\eta_n| \ll |\epsilon_n|$. 
Similarly $\eta_n$ is mass-suppressed for $n\gg n_z$. Thus mass mixing 
effects are subdominant to the pure $\epsilon_n^2$ 
corrections from direct kinetic mixing unless $m_n \sim m_Z$. 
Modes with $\Delta m = |m_n - m_Z| \ll m_Z$ 
($n \sim n_z$) can have a ``resonant enhancement'' relative to the 
kinetic mixing $\epsilon_n$:
\beq
|\eta_n| \simeq \frac{s_W\epsilon_n}{2}\lrf{m_Z}{\Delta m}
\simeq \epsilon_n\lrf{22\,\gev}{\Delta m},
\eeq
and for these modes the corrections from mass mixing effects can dominate.

  A similar discussion follows for the couplings of the KK modes $X_n$ 
to SM matter, which are given in Appendix~\ref{cov_der_app}. 
For modes with $m_n^2\ll m_Z^2$ the mass suppression of $\eta_n$ ensures 
that the dominant coupling of $X_n$ to the SM is via the electromagnetic 
current $J_{em}^\mu$. Thus the coupling relevant for low-energy probes of 
our scenario is $-c_W \epsilon_n Q_{\text em}e$. 
Similarly for $n\gg n_z$ the dominant coupling to 
the SM is via the hypercharge current $J_{Y}^\mu$.  For $n$ in the 
immediate vicinity of $n_z$ the dominant coupling 
of $X_n$ is via the SM $Z$ current $J_{Z}^\mu$, while for 
$|n-n_z|$ of order a few the coupling is via a linear 
combination of $J_{Z}^\mu$ and either $J_{EM}^\mu$ or $J_{Y}^\mu$, 
depending on the sign of $(n-n_z)$.

\section{Signals and Signatures\label{signals}}

  A warped hidden sector mixing kinetically with the SM can give rise
to new signals at the luminosity frontier achieved by fixed-target
experiments and meson factories.  Since the couplings of hidden
KK vectors to the SM are similar, it may be possible to produce multiple 
KK resonances in relatively low-energy collisions.  Thus arises the 
interesting possibility that existing and forthcoming low-energy experiments 
may directly probe an extra spacetime dimension. In this section we 
estimate the current constraints and the discovery prospects for 
this class of models.  Our analysis here is preliminary 
and a more detailed investigation will be presented in
an upcoming work~\cite{todo}.

\subsection{Hidden KK Mode Decays}

  The constraints and signals of a warped $U(1)_x$ 
gauge sector depend sensitively on how the gauge KK modes decay.  
Therefore we must determine the most likely decay channels for the 
KK excitations.  In addition to kinetic mixing with the SM on the 
UV brane, the vector KK modes also couple to the KK gravitons localized in 
the hidden bulk space.  The vector excitations can
also couple to
an explicit Higgs field on the GeV brane, 
or through higher-dimensional operators.  

  Kinetic mixing with hypercharge on the UV brane allows the gauge
KK modes to decay to the SM.  Using the results of Ref.~\cite{Batell:2009yf} 
the corresponding decay width of the
$n$-th mode to a pair of SM leptons $\ell\bar{\ell}$ is
\beq
\Gamma(X_n\to \ell\bar{\ell}) = \frac{\epsilon_n^2\,m_n}{12\pi}\,
\left(1+\frac{2m_{\ell}^2}{m_{n}^2}\right)
\left(1-\frac{m_{\ell}^2}{m_n^2}\right)^{1/2}\!\!.
\eeq
The decay width to hadrons is similar~\cite{Batell:2009yf}.
Since $\epsilon_n^2 \sim 1/n$ and $m_n \sim n$ 
the total decay width to the SM is roughly independent of mode number, 
up to a growth in the number of kinematically accessible SM final
states for heavier modes.

  In addition to  
SM decays, the heavier KK vectors can decay to 
lighter vector and graviton modes. For $n>(m+a)$ the decay 
$X_n\to X_m+h^{(a)}$ is kinematically permitted\footnote{For the Higgsed 
case this assumes $v_x\ll R^{-1}_2$ so that $X_0$ is light.} 
for $n\ge 2$ and has width
\beq
\Gamma(X_n\to X_m+h^{(a)}) \sim \frac{1}{8\pi}\lrf{k}{M_*}^2
\mathcal{B}_{a,mn}\,m_n,
\label{gravidecay}
\eeq
where $\mathcal{B}_{a,mn} \lesssim 1$ is a dimensionless coefficient
that depends on wavefunction overlaps and is suppressed if
the integers $a,\,m,\,n$ are vastly different. 
(See Appendix~\ref{app:KK_gravity} for the graviton-vector couplings).
Comparing the coupling of Eq.~\eqref{gravidecay} to Eq.~\eqref{epsilonn},
we see that gauge KK modes with
$n\geq 2$ typically decay to a
lower gauge KK mode and a graviton mode rather than going directly
to SM final states.

  If an explicit Higgs field with $U(1)_x$ charge $x_H = +1$ resides on 
the GeV brane and develops a vacuum expectation value, 
$H_x \to (v_x+\phi_x)/\sqrt{2}$ (with canonical normalization), 
the vector zero mode will acquire a mass $m_0=g_x v_x$, 
and the hidden Higgs boson $\phi_x$ will couple to the vector KK modes as
\beq
-\mathscr{L}_{eff} \supset \lambda_{mn}\,\phi_x\,X_mX_n,
\eeq
with 
\beq
\lambda_{mn} = 
\frac{g_x^2v_x}{k}\log(kR_2)\,\left.f^{(m)}\,f^{(n)}\right|_{z=R_2}.
\eeq
When kinematically allowed this coupling will induce $X_n \to \phi_x\,X_m$, 
and {will} also allow the Higgs itself to decay to a pair of 
gauge bosons. If the Higgs is lighter than the lightest gauge mode $X_0$, 
it will decay to the SM either through a loop or via 
off-shell gauge bosons as in the 4D case discussed in 
Ref.~\cite{Batell:2009yf}. While we do not consider radion 
excitations here, such a mode would participate in KK mode decays 
in much the same way as a hidden IR brane Higgs. 

Higher-dimensional operators can also contribute
to the decays of vector KK modes.  However the leading operator of this
type has the form $\int d^5x\; (X_{\mu\nu})^4/M_*^5$ 
and is typically less important than decay channels involving gravitons.  

  Putting these pieces together we see that the lightest gauge KK 
modes $X_0$ and $X_1$ 
may potentially decay primarily to pairs of SM fermions.  
For $n\geq 2$ the decays to a lighter gauge mode and a graviton, such as 
$X_2\to X_0\,h^{(1)}$, will typically dominate over the direct SM modes.  
The lightest KK graviton will also decay mainly
through the gauge-graviton coupling, 
either $h^{(1)}\to X_0X_0$ or $h^{(1)}\to X_1X_1$ 
(with $X_1$ off shell or in a loop),
and will eventually produce multiple SM fermions.
A hidden Higgs 
may also decay to the SM through gauge KK modes.

  The dominance of decays to lower KK modes rather than to the SM
coincides with the discussion of unparticles and hidden valleys
in Refs.~\cite{Strassler:2008bv,Csaki:2008dt}.  Going to higher modes, 
the decay width increases due the larger mass as well the larger number 
of final states.  This width scales as 
\beq
\Gamma_n \sim \frac{1}{8\pi}\,g_*(X_n)\lrf{k}{M_\text{Pl}}^2m_n,
\label{nwidth}
\eeq
where $g_*(X_n)$ is the number of significant decay channels. 
The results of Refs.~\cite{Strassler:2008bv,Csaki:2008dt} 
suggest that $g_*(X_n) \sim n$, 
and we will assume this to be the case here. For $k/M_\text{Pl} \ll 1$ 
the decay width is less than the mode 
separation for $n < (M_*/k)/g_*(X_n)$, while for higher modes 
the KK resonances begin to overlap. 
Thus high-energy probes initiated at the UV brane 
(such as SM initial states) will excite a continuum of overlapping
bulk modes, much like in RS2 or unparticle 
scenarios~\cite{Strassler:2008bv,Stephanov:2007ry}.\footnote{
Note that the form of the brane-to-bulk propagator
suppresses couplings to modes with $m_n \ll q$, where $q$ is the momentum
transferred in the process~\cite{ArkaniHamed:2000ds}.}  
These higher modes will cascade down to the the lightest hidden KK modes, 
which will then decay back to the SM producing a high multiplicity of 
soft final states~\cite{Csaki:2008dt}. This behavior is simply that 
of the hidden valley paradigm~\cite{Strassler:2008bv}.

\subsection{Constraints on a Hidden KK Tower}
  
  Our scenario extends the vector spectrum of Abelian hidden sector
models considered previously with an entire KK tower of new kinetically-mixed
states.  We present here a preliminary estimate of the current experimental
bounds on such a tower, deferring a detailed analysis to a future 
work~\cite{todo}.  For the time being we assume there is no
explicit Higgs-like state with mass below the lightest gauge mode
for which the bounds can be even stronger~\cite{Batell:2009di,Schuster:2009au}.
We find that the lightest gauge modes are constrained primarily by 
low-energy probes.  Bounds on heavier KK modes are less strict except 
for those with masses near the $Z$, which as discussed above can have 
a resonant enhancement in their mixing.  
We find that
low-energy probes provide the strongest constraints, and the allowed
parameter space in minimal models is very similar to that of a single 
Abelian hidden vector.

    Low-energy probes are typically the most sensitive to the 
lightest gauge mode for two reasons.  First, Eq.~\eqref{epsilonn} shows 
that the relative coupling of the higher modes to the SM is somewhat smaller, 
$\epsilon_n \sim \epsilon_0/6\sqrt{n}$.  Second, the higher modes are
heavier and can receive an additional kinematic suppression.  
For example, the constraint from the anomalous magnetic moment of 
the muon $(g\!-\!2)_{\mu}$ scales as $\epsilon_n^2/m_n^2$~\cite{Pospelov:2008zw},
and similarly for bounds from atomic parity violation~\cite{Chang:2009yw}.  
Bounds from rare meson decays are also weakened by a reduction in the final 
state phase space~\cite{Pospelov:2008zw,Reece:2009un}.  For {these} reasons, 
the 
parameter space in the $m_0\!-\!\epsilon_0$ plane for 
the full KK tower 
that is consistent with low-energy tests 
is nearly identical to that
of a single hidden Abelian gauge 
boson with $\epsilon = \epsilon_0$ and 
$m_V=m_0$~\cite{Bjorken:2009mm,Freytsis:2009bh,Pospelov:2008zw}.
With $0.3\,\gev \lesssim m_0 \lesssim 10\,\gev$, this corresponds to 
$\epsilon_0 \lesssim 3\times 10^{-3}$~\cite{Bjorken:2009mm}.\footnote{Note
that the $\epsilon$ defined in Ref.~\cite{Bjorken:2009mm} corresponds
to $\epsilon_nc_W$ here.}
 
  High-energy probes can also be sensitive to a light warped hidden 
sector.  Note that since the SM couples to the hidden sector only via
the UV brane, we can still sensibly compute processes with energies
well above the hidden IR cutoff provided they are initiated 
by SM states~\cite{ArkaniHamed:2000ds}.  Among existing higher-energy
probes, precision electroweak measurements near the $Z$ pole put the strongest
constraints on gauge kinetic mixing~\cite{Kumar:2006gm}.  
As shown in Section~\ref{sec:gauge}, 
mixing between the SM $Z$ and the hidden sector is resonantly enhanced 
for modes with $m_n\sim m_Z$.  The treatment of that section is appropriate
for narrow $X_n$ and $Z$ states.  When the mass difference 
$|m_n-m_Z|$ of an unstable gauge boson
is less than the larger of the widths $\Gamma_Z,\,\Gamma_n$,
it is more convenient to compute the effect of the hidden
tower on $e^+e^-$ collisions by treating the kinetic mixing operator
as an interaction.  

  Near the $Z$-pole the leading effect comes from two kinetic mixing
insertions on a $Z$ propagator.  This is reminiscent of an oblique
correction, but we will see that it depends in an essential way on the 
finite widths of the KK resonances.  Ignoring initial- and final-state 
fermion masses, the $Z$ propagator in the scattering amplitude is modified to
\beq
\frac{1}{(s-m_Z^2)+im_Z\Gamma_Z}
\left[1+\sum_n\frac{s_W^2\epsilon_n^2s^2}
{(s-m_Z^2+im_Z\Gamma_Z)(s-m_n^2+im_n\Gamma_n)}+\ldots
\right],
\label{zprop}
\eeq
where $\Gamma_n$ is the decay width of $X_n$.  This correction is enhanced
relative to $\epsilon_n^2$ for modes with $m_n\sim m_Z$ when $s\sim m_Z^2$,  
{coinciding} with the resonant enhancement of the mixing angle 
found 
{when} diagonalizing the full gauge boson mass matrix.  
However, we see in Eq.~\eqref{zprop} that the mixing is regulated by 
the finite decay widths.
This correction only modifies the $Z$ propagator and cancels out 
of asymmetries.  (Subleading mixing with the photon will modify 
the asymmetries.)  It can, however, modify the shape and peak location 
of the $Z$ resonance, which were measured carefully by the LEP 
collaborations~\cite{:2005ema}.  
These effects will be studied in more detail in Ref.~\cite{todo}, 
but a preliminary analysis, using Eq.~\eqref{nwidth} to estimate
the widths $\Gamma_n$, finds that the fractional shift in the lineshape
over the entire $Z$ resonance is always well below $10^{-4}$ for
$\epsilon_0 < 3\times 10^{-3}$ and KK mode spacings less than a few GeV.  
Given the precision of the combined 
electroweak measurements,
this is safely small~\cite{:2005ema}.

  Resonantly enhanced kinetic mixing can also induce non-standard 
$Z$ decays.  More precisely, the relevant process is 
$e^{+}e^{-}\to (\text{hidden})$, with the hidden sector states decaying in 
a cascade down to the lightest hidden KK modes, which subsequently decay 
to the SM. This produces a spherical distribution of multiple soft
SM final states~\cite{Csaki:2008dt}.  Indeed, this picture is precisely 
that of a hidden valley discussed in 
Refs.~\cite{Strassler:2006im,Strassler:2008bv}.
The dominant contribution comes from an $s$-channel $Z$ mixing kinetically 
into an $X_n$ state with $m_n \sim m_Z$, which then decays to lighter
graviton and gauge KK modes.  
The corresponding 
``branching fraction'', defined as a ratio of rates on the 
$Z$-pole, {is estimated} to go like
\beq
BR(Z\to\text{hidden}) \sim 
\sum_n \frac{s_W^2\epsilon_n^2m_Z^4}{(m_Z^2-m_n^2)^2+m_n^2\Gamma_n^2}\,
\lrf{\Gamma_n}{\Gamma_Z}.
\eeq
For $\epsilon_0 < 3\times 10^{-3}$ we find that this fraction 
is always less than about $3\times 10^{-6}$.  This is at the edge of 
sensitivity of the LEP experiments to exotic $Z$ decays, and the typical final
state consisting of a high multiplicity of soft leptons
and pions was not searched for directly~\cite{:2005ema,Amsler:2008zzb}.

\subsection{New Low-Energy Signals}

  We have argued that a closely-spaced tower of hidden KK modes
can be consistent with current bounds for $\epsilon_0\lesssim 3\times 10^{-3}$
and a mode spacing on the order of or below a 
GeV.  For couplings not too much smaller than this, it may be possible to 
discover the lightest hidden KK modes directly in proposed fixed-target 
and meson factory searches.  Most interestingly, several KK modes could 
potentially be found this way.

  The lowest hidden gauge mode will typically be the easiest to
find in a fixed-target experiment.  Production of this and higher modes 
will proceed as discussed in Refs.~\cite{Essig:2009nc}.  
However, relative to the
lowest mode, the production of higher modes will be suppressed
by factors of $\epsilon_n^2/\epsilon_0^2 \sim 1/36\,n$.  Despite this
suppression, the potential reach of proposed future fixed-target
experiments can exceed $\epsilon \sim 10^{-6}$~\cite{Bjorken:2009mm}, 
which may be enough to discover 
the lowest and the first excited 
hidden KK modes provided they both decay primarily to the SM.  
Note that it will 
be possible to reconstruct both resonances 
provided their mass spacing is not too small.  

  Higher resonances that decay primarily to lighter hidden states will be 
more challenging to identify in fixed-target experiments.  
On top of a reduced production rate, the methods proposed 
in Ref.~\cite{Bjorken:2009mm} rely on using specific 
geometries tailored to the distribution of signal events to reduce 
backgrounds.  In a multi-step decay process it will be more difficult 
to collect all the decay products without increasing the detector 
acceptance.  On the other hand, the multiple sets of tracks produced 
in the decay of a higher KK mode could be used to reduce backgrounds.

  Lower-energy $e^+e^-$ colliders such as BaBar, Belle, and DA$\Phi$NE
may also permit the discovery of one or more light hidden vectors.
Searches for a hidden vector based on continuum $\gamma X$ production
at these colliders currently bound $\epsilon_0 \lesssim 3\times 10^{-3}$
for $0.3\lesssim m_0 \lesssim 10\,\gev$~\cite{Bjorken:2009mm}.  This limit is
adapted from a BaBar search for $\Upsilon(3s)\to \gamma a^0$, 
where $a^0$ is a light pseudoscalar decaying to $\mu^+\mu^-$~\cite{:2009cp}.  
By expanding the search to include $\Upsilon(4s)$ data from Belle 
as well as multi-lepton final states it will be possible to investigate
a significantly larger range of hidden vector 
models~\cite{Batell:2009yf,Essig:2009nc,
Reece:2009un}, including multiple hidden vector KK modes.
Further improvements can also be expected from other low-energy searches
such as the KLOE experiment at DA$\Phi$NE~\cite{Bjorken:2009mm,Reece:2009un,Bossi:2009uw}.
    
  In addition to continuum $\gamma X$ production at $e^+e^-$ colliders, 
higher hidden KK vectors can also be produced resonantly in the 
$s$-channel for KK masses near the center-of-mass energies 
of the $B$-factories ($\sqrt{s}\sim 10.5\,\gev$) and other meson 
factories such as DA$\Phi$NE ($\sqrt{s}\sim 1.0\,\gev$)~\cite{Essig:2009nc}. 
These rates can potentially be significant when the resonant KK vectors 
are able decay efficiently into lighter hidden-sector modes.
Compared to minimal Abelian hidden scenarios where the hidden
vector is only able to decay to the SM, 
resonant production in this case is proportional to $\epsilon^2$ rather 
than $\epsilon^4$, and the heavier KK modes can be relatively broad 
making them more likely to overlap with the center-of-mass energy.  
The typical signature of a heavier vector KK mode will be multiple
charged tracks with a relatively spherical distribution, 
and can be similar to non-Abelian hidden sectors~\cite{Essig:2009nc}.
An analysis of the signals from resonant hidden KK mode
production will be be presented in Ref.~\cite{todo}.

\subsection{Dark Matter and Cosmology}

  Light hidden sectors have received attention recently in the context
of models of dark matter~(DM) motivated by new results from
PAMELA, Fermi, and DAMA.  All three of these experiments observe signals
that can potentially be due to DM.  However, 
the standard picture of a weakly-interacting massive particle 
(WIMP) undergoing thermal freeze-out does not appear to explain 
the {data}~\cite{Cirelli:2008pk}.  
Instead a DM particle with mass above a few hundred GeV coupling to 
a light hidden force with an enhanced annihilation (or decay) rate in 
the local region can 
account for the signals~\cite{ArkaniHamed:2008qn}.

  This scenario has been realized in the context of supersymmetric
hidden sectors~\cite{ArkaniHamed:2008qp}, and {with a few modifications}  
can also apply in the present scenario.
A simple option is for the DM to consist of a Dirac fermion confined to the 
UV brane with a mass near the electroweak scale and charged only
under $U(1)_x$.  This state will annihilate primarily to pairs of 
$U(1)_x$ zero-mode gauge bosons with a significant rate that depends
on the fermion mass and charge, and possibly enhanced {at late
times} through
the Sommerfeld mechanism.  These gauge bosons will in turn decay 
to SM states.  This DM candidate will also acquire a direct coupling to the 
SM $Z$ through gauge kinetic mixing, although with a strongly suppressed 
coupling.  {Let us also point out that to be consistent with
bounds from DM direct detection searches, the gauge kinetic mixing
must be somewhat small, 
$\epsilon_0 \lesssim 10^{-6}(m_0/\gev)^2$~\cite{Angle:2007uj}.}

  The primary modification required for this DM picture is in the
mechanism to generate the DM density in the early universe.
At temperatures well above the hidden IR scale the 
five-dimensional RS geometry is replaced by 
AdS-Schwarzschild~\cite{Witten:1998zw}.  
The transition from this thermal state to the usual truncated
AdS RS spacetime occurs at temperatures near the IR brane scale. 
{Depending on the mechanism of radius stabilization
this transition} can be too slow to ever complete in the calculable 
regime of $k/M_\text{Pl} \ll 1$~\cite{Creminelli:2001th}.
This possibility will be avoided 
in the present scenario
if primordial inflation never reheated above the hidden IR brane scale, 
so that the AdS-Schwarzschild geometry was never realized after inflation.  
This is consistent with primordial nucleosynthesis provided the reheating
temperature was greater than about $5\,\mev$~\cite{Hannestad:2004px}.  
With even a very low reheating temperature the DM density
could arise non-thermally 
from inflaton (or related) decays~\cite{Moroi:1999zb}.  At higher reheating 
temperatures approaching a GeV, the DM could have partially rethermalized 
enough to generate a significant relic 
abundance~\cite{Gelmini:2006pw}.

  If one is willing to give up on explaining the leptonic signals of 
PAMELA or Fermi with DM, it is also possible to have a light DM state 
within the hidden sector
that is produced thermally~\cite{Boehm:2003hm,Pospelov:2007mp}.  
Such candidates have been considered as possible explanations for 
DAMA~\cite{Gondolo:2005hh} and the INTEGRAL $511\,\kev$ 
line~\cite{Boehm:2003bt}.
A simple option is a Dirac fermion on the hidden IR brane,
but other possibilities exist. For example, instead of a minimal hidden
gauge group of $U(1)_x$ one could consider a larger hidden sector; 
one possibility would be a  warped implementation of the exact-parity 
or mirror matter models~\cite{Foot:1991bp}. If the mirror sector resides 
in the hidden bulk then the DM may be stabilized by an approximate 
hidden baryon number symmetry and additional couplings between the two 
sectors could result~\cite{Foot:2007nn}.

\section{Conclusions\label{conc}}

 In this work we have investigated a light Abelian hidden sector 
in an extended RS model. The hidden $U(1)_x$ symmetry propagates 
in a separate warped bulk and couples to the SM only via localized 
operators on a common UV brane. With a hidden IR scale of order a GeV 
the low-energy spectrum consists of the SM plus a tower of GeV 
spaced hidden KK vectors. The latter acquire a coupling to the SM via 
UV-localized mixing with SM hypercharge.  
Relative to the vector zero mode,
the couplings of the heavier KK modes are only suppressed by a moderate
logarithmic factor suggesting that they too can give rise to observable
signals.

Although we defer a detailed analysis of the bounds on such a scenario 
to a future work~\cite{todo}, a  preliminary analysis suggests that 
such a spectrum is consistent with existing constraints.
We find that the lowest KK modes can potentially decay primarily to the SM.  
Higher modes will decay in a hidden cascade down to the lowest modes, 
which then decay back to the SM.  In principle the lightest modes can 
be reconstructed directly as resonances, while the heavier modes can 
be reconstructed indirectly from their multi-body final states. 
This offers the interesting
possibility that lower-energy experiments 
operating at the luminosity frontier may observe light hidden KK vectors and 
thereby probe the structure of an extra spacetime dimension.

\textit{Note added:} 
{While this manuscript was in preparation}
Ref.~\cite{Gherghetta:2010cq} appeared, in which the authors obtain a
GeV{-scale} hidden Abelian 
{vector}
in the RS framework by an alternative means. The
phenomenology of their model differs from ours as the KK modes
of the hidden vector appear at the TeV scale in their case.
See also Ref.~\cite{Bunk:2010gb}.

\section*{Acknowledgments}

We thank Brian Batell, Yanou Cui, Damien George, Steve Godfrey, 
Tommy Levi, Rob McPherson, John Ng, Eduardo Pont\'on,
Maxim Pospelov, Adam Ritz, Steve Robertson, 
Andrew Spray, Itay Yavin and Kathryn Zurek 
for helpful comments and conversations.
This work is supported by NSERC.


\appendix
\section{Kaluza-Klein Gravitons\label{app:KK_gravity}}

  In RS models the mass of the lightest KK graviton modes is set by the 
IR scale.  In multiple-throat constructions
one expects that the KK graviton
spectrum will contain modes with masses set by the local IR scales.
In this appendix we derive the graviton KK spectrum for a pair of warped
throats sharing a common UV brane as discussed in the text.  
We confirm the
general expectation and show that this setup contains
two towers of graviton modes whose masses and splittings are set by the 
IR scales $1/R_1 \sim \tev$ and $1/R_2 \sim \gev$.  We show further that 
the $\gev$-scale modes are strongly localized in the hidden
GeV throat, and the $\tev$-scale modes are strongly localized in the 
TeV throat.  Several aspects of this discussion were
presented previously in Ref.~\cite{Cacciapaglia:2006tg}.

  The 5D graviton arises from the metric perturbation,
\begin{eqnarray}
G_{\mu\nu}^{(i)} \rightarrow (kz_i)^{-2}\left[\eta_{\mu\nu}
+\frac{2}{M_*^{3/2}}{h}_{\mu\nu}(x,z_i)\right],~~i=1,2,
\end{eqnarray}
with $\del^{\mu}h_{\mu\nu} = 
0 = h^{\mu}_{\mu}$.
Performing a KK expansion in each of the throats,
\begin{eqnarray}
h_{\mu\nu}(x,z_i)=\sum_{n}h^{(n)}_{\mu\nu}(x)f_{h,i}^{(n)}(z_i),\label{h_kk}
\end{eqnarray}
the Einstein equations require the profile in the $i$th throat to satisfy the 
equation of motion
\begin{eqnarray}
\left[z_i^2\partial_i^2-3z_i\partial_i+m_{n}^2z_i^2\right]f^{(n)}_{h,i}(z_i)=0,\label{grav_eom_f}
\end{eqnarray}
and the following orthogonality condition,
\begin{eqnarray}
\sum_i\int  \frac{dz_i}{(kz_i)^{3}}f_{h,i}^{(n)}(z_i)f_{h,i}^{(m)}(z_i)&=&\delta^{nm}.\label{kk_grav_orth}
\end{eqnarray}
The solutions to (\ref{grav_eom_f})
are of the usual RS form~\cite{Davoudiasl:1999jd},
\begin{eqnarray}
f_{h,i}^{(n)}(z_i)=\mathcal{C}^{(n)}\frac{(kz_i)^{2}}{N_{i}^{(n)}}\left\{J_{2}(m_{n}z_i)+\beta_{i}^{(n)}Y_{2}(m_{n}z_i)\right\},\label{grav_profile}
\end{eqnarray}
where $\mathcal{C}^{(n)}$ is a throat-independent 
normalization constant determined by (\ref{kk_grav_orth}), 
$\beta_{i}^{(n)}$ is a constant and we have factored out a throat-dependent 
constant $N_{i}^{(n)}$. Imposing the usual Neumann BC at the IR brane 
of the $i$th throat gives
\begin{eqnarray}
\mathrm{IR\ brane}&:& \quad\beta_{i}^{(n)}=-\frac{J_1(m_{n}R_{i})}{Y_1(m_{n}R_{i})}
\end{eqnarray}
and the demand that the metric be continuous at the common UV brane,
\begin{eqnarray}
f_{h,1}^{(n)}(k^{-1})=f_{h,2}^{(n)}(k^{-1}),
\end{eqnarray}
 determines the constants $N_{i}^{(n)}$ as
\begin{eqnarray}
N_{i}^{(n)}= J_2(m_{n}/k)+\beta_{i}^{(n)}Y_2(m_{n}/k).
\end{eqnarray}
One must also impose the generalized Israel junction condition at the UV brane~\cite{Cacciapaglia:2006tg}:
\begin{eqnarray}
\left.\sum_i \partial_i f_{h,i}^{(n)}(z_i)\right|_{UV}=0,
\end{eqnarray}
which gives
\begin{eqnarray}
\sum_i \frac{1}{N_{i}^{(n)}} \left\{J_1(m_{n}/k)+\beta^{(n)}_{i}Y_1(m_{n}/k)\right\}=0,\label{grav_kk_masses}
\end{eqnarray}
the solutions to which determine the KK masses $m_{n}$. The spectrum contains a massless mode with the throat-independent wave function
\begin{eqnarray}
f_{h,i}^{(0)}(z_i)=\sqrt{\frac{2k}{\sum_j(1-(kR_i)^{-2})}}\simeq \sqrt{k},
\end{eqnarray}
which corresponds to the usual 4D graviton. For the lighter KK modes one 
can simplify Eq.~\eqref{grav_kk_masses} by expanding in $m_n/k\ll 1$. This gives
\beq
\beta_1^{(n)}\beta_2^{(n)}=\frac{\pi}{8}\frac{m_n^2}{k^2}(\beta_1^{(n)}+\beta_2^{(n)}),
\eeq
which, for the low-lying KK masses, is well approximated by
\begin{eqnarray}
\Pi_i J_1(m_{n}R_{i}) \simeq0.\label{approx_kk_mass_grav}
\end{eqnarray}
The usual (approximate) expression for graviton KK masses in RS models 
is $J_1(m_{n}R_1)\simeq0$~\cite{Davoudiasl:1999jd} and these 
same $\sim$~TeV KK modes are contained in the spectrum (\ref{approx_kk_mass_grav}). 
In addition to these RS-like modes the spectrum contains modes with $\sim$~GeV masses 
set by $J_1(m_{h,n}R_{2})\simeq0$. Thus, the spectrum splits naturally into a pair of KK towers
with $\Delta m \sim 1/R_1 \sim \tev$ in one tower and 
$\Delta m \sim 1/R_2 \sim \gev$ in the other, 
with the lightest masses being $m_nR_{1,2} \simeq 3.83,\,7.02, ..$. 
Note that for generic $R_1,\,R_2 \ll k$, it is highly unlikely for 
both $\beta_1^{(n)}$ and $\beta_2^{(n)}$ to be simultaneously on the order
of $(m_n/k)^2\ll 1$, which would induce mixing between the towers.

  The modes in the GeV-spaced (TeV-spaced) tower are strongly localized
 in the hidden (visible) throat. As a result, the only graviton modes 
with a significant coupling to the SM are those lying in the visible throat.
This is easy to see for GeV-spaced modes considerably lighter
than a $\tev$.  The wavefunctions of these modes are approximately
flat in the visible throat and peak towards the IR brane in the GeV throat.
Computing the normalization, the KK graviton wavefunction in the TeV throat
is approximately $f_1(z_1)\simeq \sqrt{k}/kR_2$.  This amplitude is
parametrically smaller than even the zero mode graviton, 
which has $f_1(z_1) \simeq \sqrt{k}$.  The couplings of GeV-mass gravitons
to the SM can therefore be safely neglected.  This conclusion also applies
to much higher modes in the GeV tower provided $|\beta_1^{(n)}|$ is 
parametrically larger than $|\beta_2^{(n)}|$,
which is certainly the case for $m_n \lesssim \tev$.  
A similar argument applies to the TeV-spaced modes, which typically 
only have very weak couplings to the hidden sector in the GeV throat.

  The coupling between a pair of hidden KK vectors and a hidden KK graviton is given by
\beq
\mathscr{L}_{eff} \supset \frac{k}{M_\text{Pl}}\sum_{a,m,n}
\eta^{\rho\nu}\eta^{\sigma\beta}\;h_{\rho\sigma}^{(a)}\left(
{\zeta_{a,mn}}\eta^{\mu\alpha}X^m_{\mu\nu}X^n_{\alpha\beta}
-\xi_{a,mn}X^m_{\nu}X^n_{\beta}\right)
,
\label{gravicoup}
\eeq
with
\bea
\zeta_{a,mn} &=& \frac{1}{k^{3/2}}\int\!\frac{dz_2}{(kz_2)}\,
f^{(a)}_{h,2}\,f^{(m)}\,f^{(n)}\ \ ,\\
\xi_{a,mn} &=& \frac{1}{k^{3/2}}\int\!\frac{dz_2}{(kz_2)}\,
f^{(a)}_{h,2}\,\del_zf^{(m)}\,\del_zf^{(n)}\ \ ,
\eea
where $f^{(m)},f^{(n)}$ are the KK vector profiles.  

\section{Vector Mass Mixing\label{app:vec_mass_mixing}}
 The mass Lagrangian, 
\begin{eqnarray}
\mathscr{L}_{mass}&=&\frac{1}{2}m_Z^2\tilde{Z}_\mu\tilde{Z}^\mu +\frac{1}{2}\sum_{n=0}^{n_\Lambda}m_n^2\tilde{X}^n_{\mu}\tilde{X}^{\mu}_n,\nonumber
\end{eqnarray}
where $m_Z$ is the SM value of the $Z$ mass, can be written in terms of the shifted fields as 
\begin{eqnarray}
\mathscr{L}_{mass}&=&\frac{1}{2}\mathcal{V}^\mu\mathcal{M}^2\mathcal{V}_\mu,
\end{eqnarray}
where the basis vector and mass matrix are, respectively,
\begin{eqnarray}
\mathcal{V}^\mu&=&(\tilde{Z}^\mu,\tilde{X}^{\mu}_0,\tilde{X}^{\mu}_1,....,\tilde{X}^{\mu}_{n_\Lambda}),\\
\mathcal{M}^2&=&\left(\begin{array}{cc}
M_{ZZ}& M_{mix}\\
M_{mix}^T&M_{KK}\end{array}\right).
\end{eqnarray}
Here we have defined
\begin{eqnarray}
M_{ZZ}&=&m_Z^2\left(1+s_W^2\sum_{n=0}^{n_z}\epsilon_n^2\left[1+\frac{m_n^2}{m_Z^2}\right]\right),
\end{eqnarray}
and
\begin{eqnarray}
M_{KK}&=&\mathrm{diag}(\ \bar{m}_0^2\ ,\ \bar{m}_1^2\ ,....,\ \bar{m}_{n_{\Lambda}}^2\ ),
\end{eqnarray}
with 
\begin{eqnarray}
\bar{m}_n^2&\equiv&\left\{\begin{array}{lcc}(1+c_W^2\epsilon_n^2)m_n^2&\mathrm{for}&n\le n_z\\(1+\epsilon_n^2)m_n^2&\mathrm{for}&n> n_z\end{array}\right..
\end{eqnarray}
The $n$th element of the mixing vector $M_{mix}$ is given by
\begin{eqnarray}
(M_{Mix})_n&=&s_W\times\left\{\begin{array}{lll}
\epsilon_n m_n^2&\mathrm{for}& 0\le n\le n_z\\
\epsilon_n m_Z^2&\mathrm{for}& n_z<n\le n_\Lambda
\end{array}\right..
\end{eqnarray}
Diagonalization proceeds as given in the text with the mass of the physical $Z$ boson being
\begin{eqnarray}
M_{Z}^2&\simeq& m_Z^2\left(1+s_W^2\sum_{n=0}^{n_z}\epsilon_n^2\left[1+\frac{m_n^2}{m_Z^2}\right]-\sum_{n=0}^{n_{\Lambda}}\eta_n^2\left[\frac{1}{2}-\frac{m_n^2}{m_Z^2}\right]\right)\nonumber\\
& &+\sum_{n=0}^{n_{\Lambda}}\frac{2s_W^2\epsilon_n^2 m_{<}^4}{m_Z^2-m_n^2},\label{Zmass}
\end{eqnarray}
where $m_<$ is defined in (\ref{m_<}). The tower of physical hidden vectors $X_n$ have mass
\begin{eqnarray}
M_{n}^2&\simeq&  m_n^2\left(1+c_W^2\epsilon_n^2-\eta_n^2-\frac{2 s_W^2\epsilon_n^2m_n^2}{m_Z^2-m_n^2}\right)+\eta_n^2m_Z^2\quad \quad n\le n_z,\\
M_{n}^2&\simeq&  m_n^2\left(1+\epsilon_n^2-\eta_n^2\left[1-\frac{m_Z^2}{m_n^2}\right]+\frac{m_Z^2}{m_n^2}\times\frac{2 s_W^2\epsilon_n^2m_Z^2}{m_n^2-m_Z^2}\right)\quad n>n_z.
\end{eqnarray}
 With the above one readily obtains the coupling of $X_n$ to SM matter and the induced modification of the SM
 $Z$ coupling. 
\section{Gauge Boson Couplings\label{cov_der_app}}
  In the field basis where gauge kinetic mixing appears explicitly,
the couplings of the neutral gauge bosons to matter are given by
\beq
-\mathscr{L} \supset J^{\mu}_{em}A_{\mu}+J^{\mu}_ZZ_{\mu}
+\sum_nJ^{\mu}_nX_{n\,\mu}
\eeq
Shifting the fields as in Eq.~\eqref{e2_photon_shift_extended} 
and rotating them to the mass eigenbasis found above, we obtain
\bea
-\mathscr{L}&\supset& 
A_{\mu}J_{em}^{\mu}\nnmb\\ 
&&+ Z_{\mu}\left[\left(1
+\frac{1}{2}\sum_{n=0}^{n_z}s_W^2\epsilon_n^2
+\sum_{n=n_z+1}^{n_{\Lambda}}s_W\epsilon_n\eta_n
-\frac{1}{2}\sum_{n=0}^{n_{\Lambda}}\eta_n^2\right)J_Z^{\mu}\right.
\nnmb\\
&&{}~~~~~~~~~~-\left(
\sum_{n=0}^{n_z}c_Ws_W\epsilon_n^2
+\sum_{n=0}^{n_{\Lambda}}c_W\epsilon_n\eta_n
\right)
J_{em}^{\mu}
\nnmb\\
&& \left.
{}~~~~~~~~~~~~
+ \sum_{n=0}^{n_z}\left(s_W\epsilon_n+\eta_n\right)J_n
+\sum_{n=n_z+1}^{n_{\Lambda}}\eta_nJ_n\right]\\
&& + \sum_{n=0}^{n_z}X_{n\,\mu}\left[
\left(1+\frac{1}{2}c_W^2\epsilon_n^2
-s_W\epsilon_n\eta_n
-\frac{1}{2}\eta_n^2
\right)
J_n^{\mu}
-c_W\epsilon_nJ_{em}^{\mu}
-\eta_nJ_Z^{\mu}\right]
\nnmb\\
&&
+\sum_{n=n_z+1}^{n_{\Lambda}}X_{n\,\mu}\left[
\left(1+\frac{1}{2}\epsilon_n^2
-\frac{1}{2}\eta_n^2
\right)J_n^{\mu}
-c_W\epsilon_nJ_{em}^{\mu}
+\left(s_W\epsilon_n-\eta_n\right)J_Z^{\mu}\right].
\nnmb
\eea
Note that for high modes with $n\gg n_z$ (and $m_n\gg m_Z$), $\eta_n$
becomes very small so that $X_n$ couples primarily to the hypercharge
current, $J_Y = (c_WJ_{em} -s_WJ_Z)$.  Similarly, for $m_n \ll m_Z$
the corresponding state couples primarily to the electromagnetic
current.



\begin{thebibliography}{9}


\bibitem{Ball:2007zza}
  G.~L.~Bayatian {\it et al.}  [CMS Collaboration],
  J.\ Phys.\ G {\bf 34}, 995 (2007).

\bibitem{Aad:2009wy}
  G.~Aad {\it et al.}  [The ATLAS Collaboration],
  [0901.0512 [hep-ex]].

\bibitem{Morrissey:2009tf}
  D.~E.~Morrissey, T.~Plehn and T.~M.~P.~Tait,
  [0912.3259 [hep-ph]].

\bibitem{Gabbiani:1996hi}
  F.~Gabbiani, E.~Gabrielli, A.~Masiero and L.~Silvestrini,
  Nucl.\ Phys.\  B {\bf 477}, 321 (1996)
  [hep-ph/9604387].

\bibitem{Harrison:1998yr}
  P.~F.~Harrison and H.~R.~Quinn  [BABAR Collaboration],
  ``The BABAR physics book: Physics at an asymmetric $B$ factory.''

\bibitem{:2005ema}
    [ALEPH Collaboration and DELPHI Collaboration and L3 Collaboration and ],
  Phys.\ Rept.\  {\bf 427}, 257 (2006)
  [hep-ex/0509008];
LEP Electroweak Working Group,
  [0911.2604 [hep-ex]].

\bibitem{Martin:1997ns}
  S.~P.~Martin,
  [hep-ph/9709356];
  M.~A.~Luty,
  [hep-th/0509029];
  J.~Terning,
  ``Modern supersymmetry: Dynamics and duality,''
{\it  Oxford, UK: Clarendon (2006) 324 p}.

\bibitem{Nelson:1993nf}
  A.~E.~Nelson and N.~Seiberg,
  Nucl.\ Phys.\  B {\bf 416}, 46 (1994)
  [hep-ph/9309299];
  J.~Bagger, E.~Poppitz and L.~Randall,
  Nucl.\ Phys.\  B {\bf 426}, 3 (1994)
  [hep-ph/9405345];
  D.~Shih,
  JHEP {\bf 0909}, 046 (2009)
  [0906.3346 [hep-ph]];
  B.~Keren-Zur, L.~Mazzucato and Y.~Oz,
  JHEP {\bf 0909}, 041 (2009)
  [0906.5586 [hep-ph]].

\bibitem{Adriani:2008zr}
  O.~Adriani {\it et al.}  [PAMELA Collaboration],
  Nature {\bf 458}, 607 (2009)
  [0810.4995 [astro-ph]].

\bibitem{Abdo:2009zk}
  A.~A.~Abdo {\it et al.}  [The Fermi LAT Collaboration],
  Phys.\ Rev.\ Lett.\  {\bf 102}, 181101 (2009)
  [0905.0025 [astro-ph.HE]].

\bibitem{ArkaniHamed:2008qn}
  N.~Arkani-Hamed, D.~P.~Finkbeiner, T.~R.~Slatyer and N.~Weiner,
  Phys.\ Rev.\  D {\bf 79}, 015014 (2009)
  [0810.0713 [hep-ph]];
  M.~Pospelov and A.~Ritz,
  Phys.\ Lett.\  B {\bf 671}, 391 (2009)
  [0810.1502 [hep-ph]];
  Y.~Nomura and J.~Thaler,
  Phys.\ Rev.\  D {\bf 79}, 075008 (2009)
  [0810.5397 [hep-ph]];
  F.~Chen, J.~M.~Cline and A.~R.~Frey,
  Phys.\ Rev.\  D {\bf 79}, 063530 (2009)
  [0901.4327 [hep-ph]];
  F.~Chen, J.~M.~Cline and A.~R.~Frey,
  Phys.\ Rev.\  D {\bf 80}, 083516 (2009)
  [0907.4746 [hep-ph]];
  J.~T.~Ruderman and T.~Volansky,
  [0907.4373 [hep-ph]];
  J.~T.~Ruderman and T.~Volansky,
  [0908.1570 [hep-ph]].

\bibitem{Giddings:2001yu}
  S.~B.~Giddings, S.~Kachru and J.~Polchinski,
  Phys.\ Rev.\  D {\bf 66}, 106006 (2002)
  [hep-th/0105097];
  S.~Kachru, R.~Kallosh, A.~D.~Linde and S.~P.~Trivedi,
  Phys.\ Rev.\  D {\bf 68}, 046005 (2003)
  [hep-th/0301240].

\bibitem{Grana:2005jc}
  M.~Grana,
  Phys.\ Rept.\  {\bf 423}, 91 (2006)
  [hep-th/0509003];
  M.~R.~Douglas and S.~Kachru,
  Rev.\ Mod.\ Phys.\  {\bf 79}, 733 (2007)
  [hep-th/0610102];
  F.~Denef, M.~R.~Douglas and S.~Kachru,
  Ann.\ Rev.\ Nucl.\ Part.\ Sci.\  {\bf 57}, 119 (2007)
  [hep-th/0701050].


%

\bibitem{Borodatchenkova:2005ct}
  N.~Borodatchenkova, D.~Choudhury and M.~Drees,
  Phys.\ Rev.\ Lett.\  {\bf 96}, 141802 (2006)
  [hep-ph/0510147];
  S.~Heinemeyer, Y.~Kahn, M.~Schmitt and M.~Velasco,
  [0705.4056 [hep-ex]].

\bibitem{Bjorken:2009mm}
  J.~D.~Bjorken, R.~Essig, P.~Schuster and N.~Toro,
  Phys.\ Rev.\  D {\bf 80}, 075018 (2009)
  [0906.0580 [hep-ph]];
  R.~Essig, P.~Schuster, N.~Toro and B.~Wojtsekhowski,
  [1001.2557 [hep-ph]].

\bibitem{Batell:2009di}
  B.~Batell, M.~Pospelov and A.~Ritz,
  Phys.\ Rev.\  D {\bf 80}, 095024 (2009)
  [0906.5614 [hep-ph]].

\bibitem{Freytsis:2009bh}
  M.~Freytsis, G.~Ovanesyan and J.~Thaler,
  JHEP {\bf 1001}, 111 (2010)
  [0909.2862 [hep-ph]].

\bibitem{Schuster:2009au}
  P.~Schuster, N.~Toro and I.~Yavin,
  Phys.\ Rev.\  D {\bf 81}, 016002 (2010)
  [0910.1602 [hep-ph]];
  P.~Meade, S.~Nussinov, M.~Papucci and T.~Volansky,
  [0910.4160 [hep-ph]].



%
\bibitem{Pospelov:2008zw}
  M.~Pospelov,
  Phys.\ Rev.\  D {\bf 80}, 095002 (2009)
  [0811.1030 [hep-ph]].

\bibitem{Batell:2009yf}
  B.~Batell, M.~Pospelov and A.~Ritz,
  Phys.\ Rev.\  D {\bf 79}, 115008 (2009)
  [0903.0363 [hep-ph]];
  B.~Batell, M.~Pospelov and A.~Ritz,
  [0911.4938 [hep-ph]].

\bibitem{Essig:2009nc}
  R.~Essig, P.~Schuster and N.~Toro,
  Phys.\ Rev.\  D {\bf 80}, 015003 (2009)
  [0903.3941 [hep-ph]].

\bibitem{Reece:2009un}
  M.~Reece and L.~T.~Wang,
  JHEP {\bf 0907}, 051 (2009)
  [0904.1743 [hep-ph]].

\bibitem{Bossi:2009uw}
  F.~Bossi,
  [0904.3815 [hep-ex]].

\bibitem{Yin:2009mc}
  P.~f.~Yin, J.~Liu and S.~h.~Zhu,
  Phys.\ Lett.\  B {\bf 679}, 362 (2009)
  [0904.4644 [hep-ph]].

\bibitem{ArkaniHamed:2008qp}
  N.~Arkani-Hamed and N.~Weiner,
  JHEP {\bf 0812}, 104 (2008)
  [0810.0714 [hep-ph]];
  E.~J.~Chun and J.~C.~Park,
  JCAP {\bf 0902}, 026 (2009)
  [0812.0308 [hep-ph]];
  M.~Baumgart, C.~Cheung, J.~T.~Ruderman, L.~T.~Wang and I.~Yavin,
  JHEP {\bf 0904}, 014 (2009)
  [0901.0283 [hep-ph]];
  Y.~Cui, D.~E.~Morrissey, D.~Poland and L.~Randall,
  JHEP {\bf 0905}, 076 (2009)
  [0901.0557 [hep-ph]];
  C.~Cheung, J.~T.~Ruderman, L.~T.~Wang and I.~Yavin,
  Phys.\ Rev.\  D {\bf 80}, 035008 (2009)
  [0902.3246 [hep-ph]];
  A.~Katz and R.~Sundrum,
  JHEP {\bf 0906}, 003 (2009)
  [0902.3271 [hep-ph]];
  D.~E.~Morrissey, D.~Poland and K.~M.~Zurek,
  JHEP {\bf 0907}, 050 (2009)
  [0904.2567 [hep-ph]].

\bibitem{Alves:2009nf}
  D.~S.~M.~Alves, S.~R.~Behbahani, P.~Schuster and J.~G.~Wacker,
  [0903.3945 [hep-ph]];
  M.~Lisanti and J.~G.~Wacker,
  [0911.4483 [hep-ph]].

\bibitem{Strassler:2006im}
  M.~J.~Strassler and K.~M.~Zurek,
  Phys.\ Lett.\  B {\bf 651}, 374 (2007)
  [hep-ph/0604261];
  K.~M.~Zurek,
  [1001.2563 [hep-ph]].

\bibitem{Strassler:2008bv}
  M.~J.~Strassler,
  [0801.0629 [hep-ph]].

\bibitem{Georgi:2007ek}
  H.~Georgi,
  Phys.\ Rev.\ Lett.\  {\bf 98}, 221601 (2007)
  [hep-ph/0703260];
  H.~Georgi,
  Phys.\ Lett.\  B {\bf 650}, 275 (2007)
  [0704.2457 [hep-ph]].

\bibitem{Stephanov:2007ry}
  M.~A.~Stephanov,
  Phys.\ Rev.\  D {\bf 76}, 035008 (2007)
  [0705.3049 [hep-ph]];
  A.~Friedland and M.~Giannotti,
  [0709.2164 [hep-ph]];
  G.~Cacciapaglia, G.~Marandella and J.~Terning,
  JHEP {\bf 0902}, 049 (2009)
  [0804.0424 [hep-ph]];
  A.~Falkowski and M.~Perez-Victoria,
  Phys.\ Rev.\  D {\bf 79}, 035005 (2009)
  [0810.4940 [hep-ph]];
  A.~Falkowski and M.~Perez-Victoria,
  JHEP {\bf 0912}, 061 (2009)
  [0901.3777 [hep-ph]].
  A.~Friedland, M.~Giannotti and M.~Graesser,
  Phys.\ Lett.\  B {\bf 678}, 149 (2009)
  [0902.3676 [hep-th]];
  A.~Friedland, M.~Giannotti and M.~L.~Graesser,
  JHEP {\bf 0909}, 033 (2009)
  [0905.2607 [hep-th]].

\bibitem{Randall:1999ee}
  L.~Randall and R.~Sundrum,
  Phys.\ Rev.\ Lett.\  {\bf 83}, 3370 (1999)
  [hep-ph/9905221].

\bibitem{Holdom:1985ag}
  B.~Holdom,
  Phys.\ Lett.\  B {\bf 166}, 196 (1986);
  R.~Foot and X.~G.~He,
  Phys.\ Lett.\  B {\bf 267}, 509 (1991).

\bibitem{Davoudiasl:2002wz}
  H.~Davoudiasl, J.~L.~Hewett and T.~G.~Rizzo,
  JHEP {\bf 0304}, 001 (2003)
  [hep-ph/0211377];
  K.~L.~McDonald,
  Phys.\ Rev.\  D {\bf 80}, 024038 (2009)
  [0905.3006 [hep-ph]].

\bibitem{Hebecker:2006bn}
  A.~Hebecker and J.~March-Russell,
  Nucl.\ Phys.\  B {\bf 781}, 99 (2007)
  [hep-th/0607120];
  B.~v.~Harling, A.~Hebecker and T.~Noguchi,
  JHEP {\bf 0711}, 042 (2007)
  [0705.3648 [hep-th]].

\bibitem{Cacciapaglia:2006tg}
  G.~Cacciapaglia, C.~Csaki, C.~Grojean and J.~Terning,
  Phys.\ Rev.\  D {\bf 74}, 045019 (2006)
  [hep-ph/0604218].

\bibitem{ArkaniHamed:2000ds}
  N.~Arkani-Hamed, M.~Porrati and L.~Randall,
  JHEP {\bf 0108}, 017 (2001)
  [hep-th/0012148];
 R.~Rattazzi and A.~Zaffaroni,
  JHEP {\bf 0104}, 021 (2001)
  [hep-th/0012248];
  M.~Perez-Victoria,
  JHEP {\bf 0105}, 064 (2001)
  [hep-th/0105048].

\bibitem{Agashe:2007jb}
  K.~Agashe, A.~Falkowski, I.~Low and G.~Servant,
  JHEP {\bf 0804}, 027 (2008)
  [0712.2455 [hep-ph]].

\bibitem{Dimopoulos:2001ui}
  S.~Dimopoulos, S.~Kachru, N.~Kaloper, A.~E.~Lawrence and E.~Silverstein,
  Phys.\ Rev.\  D {\bf 64}, 121702 (2001)
  [hep-th/0104239];
  S.~Dimopoulos, S.~Kachru, N.~Kaloper, A.~E.~Lawrence and E.~Silverstein,
  Int.\ J.\ Mod.\ Phys.\  A {\bf 19}, 2657 (2004)
  [hep-th/0106128].

\bibitem{Cacciapaglia:2005pa}
  G.~Cacciapaglia, C.~Csaki, C.~Grojean, M.~Reece and J.~Terning,
  Phys.\ Rev.\  D {\bf 72}, 095018 (2005)
  [hep-ph/0505001].

\bibitem{Flacke:2006ad}
  T.~Flacke, B.~Gripaios, J.~March-Russell and D.~Maybury,
  JHEP {\bf 0701}, 061 (2007)
  [hep-ph/0611278];
  T.~Flacke and D.~Maybury,
  JHEP {\bf 0703}, 007 (2007)
  [hep-ph/0612126];
  B.~Gripaios,
  Nucl.\ Phys.\  B {\bf 768}, 157 (2007)
  [hep-ph/0611218];
  B.~v.~Harling and A.~Hebecker,
  JHEP {\bf 0805}, 031 (2008)
  [0801.4015 [hep-ph]];
  A.~Bechinger and G.~Seidl,
  [0907.4341 [hep-ph]].

\bibitem{Batell:2005wa}
  B.~Batell and T.~Gherghetta,
  Phys.\ Rev.\  D {\bf 73}, 045016 (2006)
  [hep-ph/0512356].

\bibitem{Dienes:1996zr}
  K.~R.~Dienes, C.~F.~Kolda and J.~March-Russell,
  Nucl.\ Phys.\  B {\bf 492}, 104 (1997)
  [hep-ph/9610479].

\bibitem{Abel:2003ue}
  S.~A.~Abel and B.~W.~Schofield,
  Nucl.\ Phys.\  B {\bf 685}, 150 (2004)
  [hep-th/0311051];
  S.~A.~Abel, M.~D.~Goodsell, J.~Jaeckel, V.~V.~Khoze and A.~Ringwald,
  JHEP {\bf 0807}, 124 (2008)
  [0803.1449 [hep-ph]];
  M.~Goodsell, J.~Jaeckel, J.~Redondo and A.~Ringwald,
  JHEP {\bf 0911}, 027 (2009)
  [0909.0515 [hep-ph]];
  M.~Goodsell and A.~Ringwald,
  [1002.1840 [hep-th]].

\bibitem{Agashe:2003zs}
  K.~Agashe, A.~Delgado, M.~J.~May and R.~Sundrum,
  JHEP {\bf 0308}, 050 (2003)
  [hep-ph/0308036];
  K.~Agashe, R.~Contino, L.~Da Rold and A.~Pomarol,
  Phys.\ Lett.\  B {\bf 641}, 62 (2006)
  [hep-ph/0605341].

\bibitem{Carena:2006bn}
  M.~S.~Carena, E.~Ponton, J.~Santiago and C.~E.~M.~Wagner,
  Nucl.\ Phys.\  B {\bf 759}, 202 (2006)
  [hep-ph/0607106];
  M.~S.~Carena, E.~Ponton, J.~Santiago and C.~E.~M.~Wagner,
  Phys.\ Rev.\  D {\bf 76}, 035006 (2007)
  [hep-ph/0701055].

\bibitem{Csaki:2003dt}
  C.~Csaki, C.~Grojean, H.~Murayama, L.~Pilo and J.~Terning,
  Phys.\ Rev.\  D {\bf 69}, 055006 (2004)
  [hep-ph/0305237];
  C.~Csaki, C.~Grojean, L.~Pilo and J.~Terning,
  Phys.\ Rev.\ Lett.\  {\bf 92}, 101802 (2004)
  [hep-ph/0308038].

\bibitem{Randall:1999vf}
  L.~Randall and R.~Sundrum,
  Phys.\ Rev.\ Lett.\  {\bf 83}, 4690 (1999)
  [hep-th/9906064].

\bibitem{Goldberger:1999uk}
  W.~D.~Goldberger and M.~B.~Wise,
  Phys.\ Rev.\ Lett.\  {\bf 83}, 4922 (1999)
  [hep-ph/9907447].

\bibitem{Csaki:2008zd}
  C.~Csaki, A.~Falkowski and A.~Weiler,
  JHEP {\bf 0809}, 008 (2008)
  [0804.1954 [hep-ph]].

\bibitem{Pomarol:2000hp}
  A.~Pomarol,
  Phys.\ Rev.\ Lett.\  {\bf 85}, 4004 (2000)
  [hep-ph/0005293].

\bibitem{todo}
K.~L.~McDonald and D.~E.~Morrissey, in preparation.

\bibitem{Csaki:2008dt}
  C.~Csaki, M.~Reece and J.~Terning,
  JHEP {\bf 0905}, 067 (2009)
  [0811.3001 [hep-ph]].

\bibitem{Chang:2009yw}
  W.~F.~Chang, J.~N.~Ng and J.~M.~S.~Wu,
  Phys.\ Rev.\  D {\bf 79}, 055016 (2009)
  [0901.0613 [hep-ph]].

\bibitem{Kumar:2006gm}
  J.~Kumar and J.~D.~Wells,
  Phys.\ Rev.\  D {\bf 74}, 115017 (2006)
  [hep-ph/0606183];
  W.~F.~Chang, J.~N.~Ng and J.~M.~S.~Wu,
  Phys.\ Rev.\  D {\bf 74}, 095005 (2006)
  [Erratum-ibid.\  D {\bf 79}, 039902 (2009)]
  [hep-ph/0608068];
  D.~Feldman, Z.~Liu and P.~Nath,
  Phys.\ Rev.\  D {\bf 75}, 115001 (2007)
  [hep-ph/0702123].

\bibitem{Amsler:2008zzb}
  C.~Amsler {\it et al.}  [Particle Data Group],
  Phys.\ Lett.\  B {\bf 667}, 1 (2008).

\bibitem{:2009cp}
  B.~Aubert {\it et al.}  [BABAR Collaboration],
  [0902.2176 [hep-ex]];
  B.~Aubert {\it et al.}  [BABAR Collaboration],
  Phys.\ Rev.\ Lett.\  {\bf 103}, 081803 (2009)
  [0905.4539 [hep-ex]].


\bibitem{Cirelli:2008pk}
  M.~Cirelli, M.~Kadastik, M.~Raidal and A.~Strumia,
  Nucl.\ Phys.\  B {\bf 813}, 1 (2009)
  [0809.2409 [hep-ph]].

\bibitem{Angle:2007uj}
  J.~Angle {\it et al.}  [XENON Collaboration],
  Phys.\ Rev.\ Lett.\  {\bf 100}, 021303 (2008)
  [0706.0039 [astro-ph]];
  Z.~Ahmed {\it et al.}  [The CDMS-II Collaboration],
  [0912.3592 [astro-ph]].

\bibitem{Witten:1998zw}
  E.~Witten,
  Adv.\ Theor.\ Math.\ Phys.\  {\bf 2}, 505 (1998)
  [hep-th/9803131].

\bibitem{Creminelli:2001th}
  P.~Creminelli, A.~Nicolis and R.~Rattazzi,
  JHEP {\bf 0203}, 051 (2002)
  [hep-th/0107141];
  L.~Randall and G.~Servant,
  JHEP {\bf 0705}, 054 (2007)
  [hep-ph/0607158];
  J.~Kaplan, P.~C.~Schuster and N.~Toro,
  [hep-ph/0609012];
  B.~Hassanain, J.~March-Russell and M.~Schvellinger,
  JHEP {\bf 0710}, 089 (2007)
  [0708.2060 [hep-th]].

\bibitem{Hannestad:2004px}
  S.~Hannestad,
  Phys.\ Rev.\  D {\bf 70}, 043506 (2004)
  [astro-ph/0403291].

\bibitem{Moroi:1999zb}
  T.~Moroi and L.~Randall,
  Nucl.\ Phys.\  B {\bf 570}, 455 (2000)
  [hep-ph/9906527].

\bibitem{Gelmini:2006pw}
  G.~B.~Gelmini and P.~Gondolo,
  Phys.\ Rev.\  D {\bf 74}, 023510 (2006)
  [hep-ph/0602230].

\bibitem{Boehm:2003hm}
  C.~Boehm and P.~Fayet,
  Nucl.\ Phys.\  B {\bf 683}, 219 (2004)
  [hep-ph/0305261];
  C.~Boehm, P.~Fayet and J.~Silk,
  Phys.\ Rev.\  D {\bf 69}, 101302 (2004)
  [hep-ph/0311143];
  P.~Fayet,
  Phys.\ Rev.\  D {\bf 70}, 023514 (2004)
  [hep-ph/0403226].

\bibitem{Pospelov:2007mp}
  M.~Pospelov, A.~Ritz and M.~B.~Voloshin,
  Phys.\ Lett.\  B {\bf 662}, 53 (2008)
  [0711.4866 [hep-ph]];
  D.~Hooper and K.~M.~Zurek,
  Phys.\ Rev.\  D {\bf 77}, 087302 (2008)
  [0801.3686 [hep-ph]];
  J.~L.~Feng and J.~Kumar,
  Phys.\ Rev.\ Lett.\  {\bf 101}, 231301 (2008)
  [0803.4196 [hep-ph]];
  K.~M.~Zurek,
  Phys.\ Rev.\  D {\bf 79}, 115002 (2009)
  [0811.4429 [hep-ph]].

\bibitem{Gondolo:2005hh}
  P.~Gondolo and G.~Gelmini,
  Phys.\ Rev.\  D {\bf 71}, 123520 (2005)
  [hep-ph/0504010];
  F.~Petriello and K.~M.~Zurek,
  JHEP {\bf 0809}, 047 (2008)
  [0806.3989 [hep-ph]];
  C.~Savage, G.~Gelmini, P.~Gondolo and K.~Freese,
  JCAP {\bf 0904}, 010 (2009)
  [0808.3607 [astro-ph]].

\bibitem{Boehm:2003bt}
  C.~Boehm, D.~Hooper, J.~Silk, M.~Casse and J.~Paul,
  Phys.\ Rev.\ Lett.\  {\bf 92}, 101301 (2004)
  [astro-ph/0309686].

\bibitem{Foot:1991bp}
  R.~Foot, H.~Lew and R.~R.~Volkas,
  Phys.\ Lett.\  B {\bf 272}, 67 (1991).

\bibitem{Foot:2007nn}
  R.~Foot,
  Int.\ J.\ Mod.\ Phys.\  A {\bf 22}, 4951 (2007)
  [0706.2694 [hep-ph]].

\bibitem{Gherghetta:2010cq}
  T.~Gherghetta and B.~von Harling,
  [1002.2967 [hep-ph]].

\bibitem{Bunk:2010gb}
  D.~Bunk and J.~Hubisz,
  [1002.3160 [hep-ph]].


\bibitem{Davoudiasl:1999jd}
  H.~Davoudiasl, J.~L.~Hewett and T.~G.~Rizzo,
  Phys.\ Rev.\ Lett.\  {\bf 84}, 2080 (2000)
  [hep-ph/9909255].




\end{thebibliography}
\end{document}